\documentclass{article}
\usepackage{dcolumn}
\usepackage{bm}
\usepackage{verbatim}       

\usepackage[dvips]{graphicx}
\usepackage{amsthm}
\usepackage{bm}
\usepackage{amstext}
\usepackage{amsmath}
\usepackage{amsfonts}

\newcommand{\dd}{{\rm d}}
\newcommand{\bd}{\begin{definition}}                
\newcommand{\ed}{\end{definition}}                  
\newcommand{\bc}{\begin{corollary}}                 
\newcommand{\ec}{\end{corollary}}                   
\newcommand{\bl}{\begin{lemma}}                     
\newcommand{\el}{\end{lemma}}                       
\newcommand{\bp}{\begin{proposition}}            
\newcommand{\ep}{\end{proposition}}                
\newcommand{\bere}{\begin{remark}}                  
\newcommand{\ere}{\end{remark}}                     

\newcommand{\bt}{\begin{theorem}}
\newcommand{\et}{\end{theorem}}

\newcommand{\be}{\begin{equation}}
\newcommand{\ee}{\end{equation}}

\newcommand{\bit}{\begin{itemize}}
\newcommand{\eit}{\end{itemize}}
\newtheorem{theorem}{Theorem}[section]
\newtheorem{corollary}[theorem]{Corollary}
\newtheorem{lemma}[theorem]{Lemma}
\newtheorem{proposition}[theorem]{Proposition}
\theoremstyle{definition}
\newtheorem{definition}[theorem]{Definition}
\theoremstyle{remark}
\newtheorem{remark}[theorem]{Remark}

\begin{document}
%

\title{On the global existence of time\thanks{Slightly improved version of the essay awarded the third juried prize at the FQXi contest ``The Nature of Time".}}

\author{E. Minguzzi\thanks{
Dipartimento di Matematica Applicata ``G. Sansone'', Universit\`a
degli Studi di Firenze, Via S. Marta 3,  I-50139 Firenze, Italy.
E-mail: ettore.minguzzi@unifi.it} }

\date{}


\maketitle

\begin{abstract}
\noindent The existence of a global  time is often taken for granted
but should instead be considered as a matter of investigation. By
using the tools of global Lorentzian geometry  I show that, under
physically reasonable conditions, the impossibility of finding a
global time implies the singularity of  spacetime.
\end{abstract}

\section{Introduction}

Many questions on the nature of time tacitly assume that a global
time does indeed exists, that a time, perhaps associated to the
cosmological flow, makes sense all over the spacetime. If we put the
problem of the global existence of time under scrutiny we soon
realize that it has no simple answer. Indeed,  the properties of
time at very large scales should be inferred from local observations
which, however, are limited.

With the discovery of the microwave background radiation we have
learned that the `deepest' photons that reach us come from the last
scattering surface, a region of space of huge extension that
nevertheless is not expected to embrace the whole Universe. Is it
therefore hopeless to try to justify the global existence of time?
Can we speak of a global time, or should we resign and treat time
only as a local phenomenon?

I wish to prove that a lot can be learned on the global nature of
time, indeed I shall draw a connection between the existence of time
as a global entity and the absence of singularities in the Universe.
The possibility of obtaining such connection, apparently paradoxical
given our limited observational capabilities, stands on the fact
that although our observations are local, the laws that we find hold
globally (a physical principle that has passed several tests, think
of the proportions between the absorption lines of chemical
elements, which are the same on earth as well as in distant
galaxies). These local laws can therefore put global constraints on
spacetime and thus say something on the existence of a global time.
As we shall see, these results will be obtained by using the tools
of global Lorentzian geometry, that is, that branch of mathematics
which studies the global aspects of Einstein's general theory of
relativity.

\section{Times}

Before we proceed it is useful to distinguish between two different
concepts of time which are used in physics and in particular in
general relativity. They both deserve to be called {\em time}, as
they are actually the same in a non-relativistic (low velocity)
context. However, without a clarification some confusion could arise
because they differ considerably outside that limit.

The first notion of time is the one that retains the property of
{\em reckoning}. We are used to the fact that time is measured with
clocks, more precisely clocks are used to associate to any pair of
successive events a dimensional number called {\em interval of
time}. This aspect of time is often called the {\em metric} or the
{\em distance} side of time. It is intimately connected with the
problem of time measurement which in turn influences the very
definition of this type of time concept. This notion of time in
general relativity becomes path dependent, the so called {\em proper
time}: it is meaningful to take intervals of time only along the
path of an observer, mathematically represented as a timelike curve.
It cannot be exported to a global entity because of the well known
clock effect (twin paradox): different observers would disagree on
the label to assign to a given event since different observers have
in general different histories and hence a different final reading
of the respective clocks.

Fortunately, time is not only reckoning, but has an even more
fundamental role, and here we come to the other side of time, that
of {\em order}. Clearly, in non-relativistic physics time can be
used to order events. If events $a$ and $b$ are such that
$t(b)-t(a)>0$ then we say that $b$ has happened {\em after} $a$. We
see that here the order role is deduced from the reckoning role,
because the order follows from the sign of the time interval
$t(b)-t(a)$ (if it vanishes the events  are said to be
simultaneous). If we think about it we can easily realize that in
non-relativistic physics the time order has a {\em causal} nature.
If event $a$ can influence event $b$ then $t(b)
>t(a)$.
By passing to the relativistic context  we want the new time to be
compatible with the more complex causal structure of spacetime as it
follows from the finiteness of the speed of light (i.e.\ from the
existence of the light cones) and from the curvature of spacetime.
Physicists have learned how to express this notion of time, which is
compatible with causality and retains an ordering role; they call it
{\em the time function}.

Before we give a more precise definition let use recall some basics
of relativistic physics. The spacetime of general relativity,
denoted $(M,g)$, is  a 4-dimensional manifold whose elements are
called {\em events}. This manifold is endowed with a Lorentzian
metric, that is, a point dependent metric of signature $(-,+,+,+)$.
At a given event tangent vectors $v$ separate into timelike,
lightlike and spacelike depending on the value of $g(v,v)$,
respectively negative, zero or positive. Lightlike or timelike
vectors are called causal, and the terminology extends to curves
depending on the character of their tangent vectors. Note that in
the tangent space of every point there are two cones of timelike
vectors. It is assumed that at each point one of the cones is
labeled {\em future} and the other {\em past} and that this labeling
is actually continuous over the manifold (i.e. the Lorentzian
manifold is time oriented).


%

Mathematically, in general relativity event $a$ can influence event
$b$ if there is a causal curve connecting the two events. The idea
is that signals propagate on causal curves and thus their tangent
vector must be contained in the future light cone. Following a
customary notation, we shall write $a <b$, and call $<$ the {\em
causality relation}. If $a<b$ or $a=b$ we write $a\le b$, and if
there is a timelike curve connecting them we write $a\ll b$.
Observers and massive particles are represented by timelike curves,
light beams by lightlike geodesics.

Given these definitions it is now easy to define what is a time
function

\begin{definition}
A {\em time function} is a continuous function $t: M \to \mathbb{R}$
such that if $a<b$ then $t(a)<t(b)$.
\end{definition}

It is the statement ``$a<b \Rightarrow t(a)<t(b)$'' which conveys in
mathematical terms the compatibility of  time with the causal
structure. It states that if $a$ can influence $b$ then the time of
$a$ must be less than that of $b$. However, a time $t$ must not
necessarily exist while the concept of causal influence is more
primitive and built in the very definition of spacetime. There are
also other properties that deserve attention in the definition of
time function. First, the domain of $t$ is the whole $M$ which means
that it is global; second, it is continuous, and this appears as a
minimal mathematical and physical requirement. Thus in rigorous
terms, speaking of a global time means speaking of a time function.

Note that in a spacetime there may be pairs of events $(a,b)$,  for
which neither $a\le b$ nor $b\le a$ hold, in contrast with the
non-relativistic context. Let us recall that a relation $R$ is {\em
total} if $aRb$ or $bRa$. Moreover, $R$ is a preorder of it is
reflexive (i.e. $aRa$) and transitive, while it is a partial order
if it is an antisymmetric preorder (antisymmetry: $aRb$ and $bRa$
$\Rightarrow a=b$). Thus, stated in another way, the relation $\le $
could be a partial order (this property is called causality and is
equivalent to the absence of closed causal curves) but in general it
is not a total preorder, that is it cannot ``decide'' for all pairs
which event comes before and which after. Here the ordering role of
time comes into play. If you have a time function you also have a
new total preorder, indeed you can establish that $b$ {\em comes
after or at the same time of} $a$ if $t(a)\le t(b)$. The new
relation is indeed a total preorder which is an extension of the
causal relation ``$\le$''.

In general if a spacetime admits a time function then this function
needs not to be unique. This feature is characteristic of
relativity, indeed already in special relativity each observer has
its own time function. The problem of building a correspondence
between observers and time functions is  quite interesting but will
not be addressed in this work \cite{minguzzi03}. To a large extent
it is in fact a problem of methodology, it involves the problem of
synchronization of clocks and its limitations in a general
relativistic framework \cite{minguzzi02d}. However, it is not a
fundamental issue on the nature of time although its study clarifies
the actual methods through which observers build up a time
coordinate on portions of spacetime. Instead, here we ask a more
fundamental issue, because maybe the spacetime does not even admit
{\em one} time function.

\section{Time functions and causality}

Does a time function always exist? This is a rephrasing of our
original question. In mathematical terms the answer is negative
because it is easy to construct spacetimes that do not comply this
condition. Consider for instance Minkowski spacetime with the usual
coordinates and the space slices $t=-1$ and $t=+1$ identified. The
spacetime has a toroidal shape and admits closed timelike curves
(e.g. the curve of equation $x^i=0$, $i=1,2,3$). It is clear that
any spacetime admitting a closed causal curve cannot admit a time
function. Indeed, the function would have to increase all over the
curve, which is impossible because the final endpoint would coincide
with the starting one. Thus arbitrary spacetimes need not admit  a
time function.

The reader could suggest that perhaps the just mentioned spacetime
is not realistic and should be discarded. This criticism is not
particularly convincing because this spacetime actually satisfies
the Einstein equations. It turns out that in the context of general
relativity it is unnatural  to dismiss {\em a priori} spacetimes,
like this one, which present closed timelike curves. The reason why
these spacetime are not generally accepted is mostly a philosophical
one. A closed timelike curve represents an observer which is forced
to live an infinite number of times the same history. It is a kind
of backward time travel in which the free will of the observer seems
unable to affect the physical evolution (the grandfather paradox).
Most physicists take therefore the view that the spacetimes which
present this chronology violation should not be regarded as
physical. Other physicists claim that although that conclusion could
be correct, its validity  should be a matter of investigation
\cite{thorne93}. They claim that perhaps introducing the effects of
quantum mechanics it could be possible to prove that spacetimes
presenting chronology violations would evolve shrinking those
causality violating regions, so that in the end they would have
little influence on the global aspects of spacetime. In this work I
shall consider the spacetimes as free from closed timelike curves, a
minimal  and philosophically satisfactory assumption that is known
as the {\em chronology condition}.

\begin{figure}[ht]
\begin{center}
 \includegraphics[width=9cm]{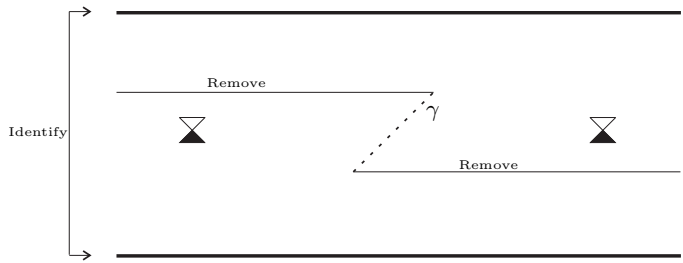}
\end{center}
\caption{A chronological and causal spacetime that does not admits a
time function. Past cones are depicted in black.} \label{p1}
\end{figure}

Unfortunately, even chronological or causal spacetimes may not admit
a time function. In order to show  this fact I present the classical
\cite{hawking70} example of figure \ref{p1}.

Here, as well as in other parts of the work, the example is obtained
from 1+1 Minkowski spacetime by removing some parts and by making
some identifications. The reader could be concerned that the
spacetimes so obtained are not `realistic'; in this respect some
observations are in order. These spacetimes are intended only to
clarify the logical independence of some concepts; the restricted
dimensionality is not important as by multiplication with a $S^2$
sphere one would get 4-dimensional examples; moreover, although they
seem part of a larger Minkowski spacetime (i.e.\ they seem to miss
something), this is not really so. Indeed, as proved by Beem
\cite{beem76}, one can always use the trick of multiplying the flat
metric by a conformal factor so that the spacetimes become causally
geodesically complete (every causal geodesic has an affine parameter
which takes all the values of $\mathbb{R}$). After this operation
the spacetime cannot be further extended (as corresponding to an
extension one would expect the geodesics to extend to the new region
and so the affine parameters of the geodesics to extend too which is
impossible) and thus there are no missing points as it could be
suggested by the original operation that led to their construction.

Returning to the example of figure \ref{p1} we see that no closed
causal curve exists as no causal curve can cross the dotted line.
The spacetime is therefore causal. The reader can try to construct a
time function on this spacetime and check that despite the effort it
is impossible. There is indeed a beautiful theorem by S. Hawking
\cite{hawking68} which states that a spacetime admits a time
function if and only if it is stably causal. This theorem was
subsequently refined to show that the time function, whenever it
exists,  can always be chosen to be smooth with a timelike gradient
\cite{bernal04}. Stable causality is a condition which is stronger
than causality (which in turn is stronger than chronology) and
states that not only there must not be closed causal curves on
spacetime but also that it is possible to slightly open up the light
cones all over the spacetime (thus changing the causal structure)
without introducing closed causal curves. In the example of figure
\ref{p1} opening the light cones means to allow the causal curves to
tilt slightly more than the dotted line $\gamma$, a fact which
allows to construct a closed causal curve. Therefore, the spacetime
of figure \ref{p1} is  not stably causal.

The reader may still wonder if there are spacetimes without the
strange corners and removed sets of figure \ref{p1} which are not
stably causal. Again the answer is affirmative. The spacetime of
topology $\mathbb{R}^4$, coordinates $q_1,q_2,u,y$ and metric (here
$r=\sqrt{q_1^2+q_2^2}$)
\begin{equation}
\dd s^2=\dd q_1^2+\dd q_2^2 \!-\!\dd u \otimes [\dd y-r(q_1\dd
q_2-q_2\dd q_1)] -\![\dd y-\!r(q_1\dd q_2-q_2\dd q_1)] \otimes \dd
u,
\end{equation}
does not admit a time function (even more it is non-future
distinguishing, that is there are different points with the same
chronological future; for another example with a nice figure see
\cite{hubeny05}).

\begin{figure}[ht]
\begin{center}
 \includegraphics[width=9cm]{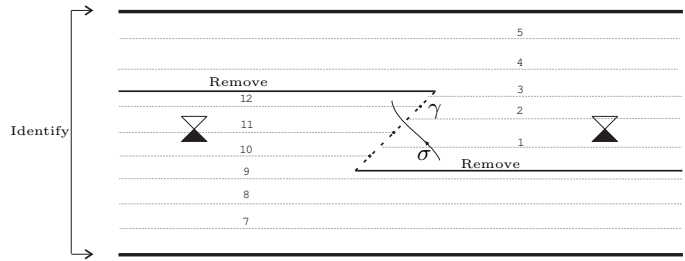}
\end{center}
\caption{The constant slices of a function which increases on every
causal curve. For instance it increases over the timelike curve
$\sigma$, but note that this function is discontinuous all over the
lightlike geodesic $\gamma$. } \label{p0}
\end{figure}

According to our analysis and Hawking's theorem we have to prove
starting from chronology that the spacetime is stably causal, a fact
that the previous examples prove to be false unless we add some
additional requirement. The right suggestion here comes from the
mathematical field known as the ``theory of relations''. A general
fact proved by Szpilrajn (1930) about partial orders (i.e.\
reflexive, transitive binary relations that satisfy the antisymmetry
property $(x,z)\in R$ and $(z,x) \in R \Rightarrow x=z$), is that a
partial order  can be extended to a total order \cite[Sect.
2.9.3]{fraisse00} and hence to a total preorder. Once applied to the
causal relation $\le$, taking into account that a time function
provides a total preorder, this result seems to suggest that
causality implies the existence of a total preorder and thus maybe
also the existence of a time function. We know that this is not
possible, a tricky situation which can be understood if we look at
figure \ref{p0}.

Here I have drawn the constant slices of a function which indeed
increases on every causal curve. This function does indeed provide a
total preorder for the spacetime events but it is not continuous!
The continuity property is fundamental in Hawking's theorem and in
the very definition of time function. Thus Hawking's and Szpilrajn's
results are actually compatible. The discussion suggests that
perhaps if we could prove that there are no discontinuity points
then we could also prove the existence of a time function. Note that
in the figure the discontinuity points lie in the dotted lightlike
geodesic $\gamma$. This causal curve has some further properties: it
cannot be further extended neither in the past nor in the future,
that is, it is inextendible, and  no two points of the curve can be
joined by a timelike curve. In other words, this curve is what
physicists call a {\em lightlike line}. The idea is then to add to
causality the property of ``absence of lightlike lines'' with the
hope that with this assumption one could remove all the
discontinuity points and thus prove the existence of a time
function. Actually, it is a trivial result that chronology plus the
absence of lightlike lines implies causality \cite{hawking73} thus
the statement to be proved
is\\

\noindent {\bf Theorem}. Chronological spacetimes without lightlike
lines are stably causal.\\

Remarkably, recently I gave two largely independent proofs of this
fact \cite{minguzzi07d,minguzzi08b}. The second proof also solves a
long standing conjecture in causality theory, that on the possible
equivalence between stable causality and $K$-causality
\cite{sorkin96} (see \cite{minguzzi07} for an historical
perspective). The theorem is interesting in its own right as it has
no Riemannian analog and involves only the light cone structure of
spacetime (i.e. the reckoning aspect of time does not enter).

I shall not comment the details of the proof here. The point is,
does this result say something on the existence of a global time?
After all we have reduced the existence of time to that of ``absence
of lightlike lines'' (as we accepted chronology), but is this a
progress? This question is answered in the next section.

\section{The positivity of the energy density}.

The absence of lightlike lines is implied by some physically
reasonable conditions. This fact was first used by Hawking and
Penrose in the proof of their singularity theorem \cite{hawking70}.

Assume that the spacetime is
\begin{itemize}
\item[(i)] Null geodesically complete,
\end{itemize}
that is every inextendible lightlike geodesic $x(\lambda)$ has an
affine parameter $\lambda$ which runs from $-\infty$ to $+\infty$.
This condition means that the spacetime does not end abruptly, it is
therefore regarded as a non-singularity requirement. Actually,
physicists consider other non-singularity requirements, for instance
that which demands  the spacetime to be timelike geodesically
complete. These requirements are logically independent as a
spacetime can be, for instance, timelike geodesically incomplete but
null geodesically complete \cite{beem96}. Condition (i) must be read
as a weak non-singularity requirement.

The second assumption is that the spacetime satisfies the null
convergence condition
\begin{itemize}
\item[(ii)] $R_{\mu \nu}n^{\mu}n^{\nu}\ge 0$ for all lightlike
vectors $n^{\mu}$,
\end{itemize}
where $R_{\mu \nu}$ is the Ricci tensor. Recall that the Einstein
equations are
\[
R_{\mu \nu}-\frac{1}{2}Rg_{\mu \nu}+\Lambda g_{\mu \nu}=8 \pi G
T_{\mu \nu}.
\]
If the energy density for an observer of 4-velocity $u^{\mu}$ is
non-negative then $T_{\mu \nu} u^{\mu} u^{\nu}\ge 0$, and since this
is a reasonable assumption (in a non-quantum mechanical regime) then
this inequality should hold for every timelike vector $u^{\mu}$ and
hence, by continuity, for every lightlike vector $n^{\mu}$. Using
the Einstein equations  and $T_{\mu \nu} n^{\mu} n^{\nu}\ge 0$ we
get the assumption (ii) which can therefore be regarded as a
consequence of the positivity of the energy density. Actually this
assumption can also be weakened to the {\em averaged null
convergence condition} \cite{tipler78,tipler78b,chicone80} which
allows for local violations of the positive energy condition.

The last assumption is the most technical one and is called the {\em
null genericity condition}
\begin{itemize}
\item[(iii)] Every inextendible  lightlike geodesic $\gamma$  has at some point a  tangent vector $n$ such that, $n^{c} n^{d} n_{[a} R_{b] c d [e} n_{f]} \ne 0$.
\end{itemize}
Basically it states that the spacetime is not too `special' in the
sense that it does not have particular symmetries. This condition is
physically reasonable because if violated it could be restored
through an arbitrarily small perturbation of the metric along the
lightlike geodesic.

A well known result by Hawking and Penrose \cite{hawking70} states
that if (i), (ii) and (iii) hold then the spacetime does not have
lightlike lines.  Mathematicians have also studied what happens if
(i) and (ii) hold but there are lightlike lines (because (iii)
fails). The results is that, as expected from the failure of (iii),
the spacetime would have rather special features \cite[Theorem
IV.1]{galloway00}.

We can understand this result by Hawking and Penrose with a
Riemannian analogy. Imagine you live in a world with two space
dimensions, which we  model with a Riemannian manifold, and an
absolute time. In this model you are not a good guy so you steal the
wallet of somebody and run away. Unfortunately, you choose the wrong
person so that you are immediately chased by a multitude of people
running  after you. They run exactly at your speed so you are forced
to move along a space geodesic. Now, if the space were flat you
would be able to keep the same distance between you and the people
following you.

\begin{figure}[ht]
\begin{center}
 \includegraphics[width=7cm]{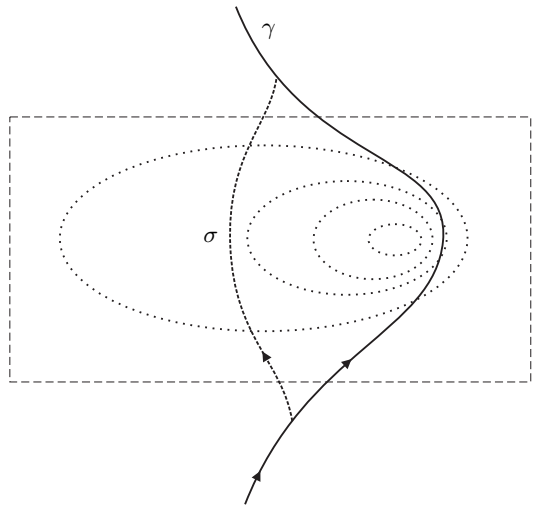}
\end{center}
\caption{Your path and that of most of the people following you is
the geodesic $\gamma$. The minimization property of geodesics holds
locally but not necessarily globally if the space is curved. Someone
of the group behind you can take advantage of the presence of the
hill to follow path $\sigma$ and catch up with you. This is possible
if the space and hence the geodesics extends enough. If the whole
space were only the portion inside the box then the geodesics would
be incomplete (the space would be singular) and the followers would
not catch you.} \label{haw}
\end{figure}

If the space is curved, however, some persons in the group behind
you can choose to try a different path and surprise you. This is
possible because geodesics are the paths that {\em locally} minimize
the length between two points. However, globally the geodesics may
lose this property. Figure \ref{haw} shows what could happen if in
your running away you pass nearby a hill. Provided the geodesics can
be extended sufficiently far this strategy will work and your
followers will catch you. The energy and the genericity conditions
(ii)-(iii) basically state that the universe is curved, so that the
analog of the hill exists.

Putting this result together with that of the previous section, and
assuming chronology plus the physically reasonable conditions (ii)
and (iii), we get that {\em under physically reasonable conditions
if the universe is null geodesically complete then there is a time
function}. Stated in another way, {\em under physically reasonable
conditions if a time function does not exist then there is a
singularity}. Note that this singularity theorem does not assume the
existence of a time function from start, as  Penrose's (1965) does
\cite{hawking73}, nor it does assume the existence of a closed
trapped surface (i.e.\ a surface that traps light). Indeed, it is in
a sense a more primitive result, not directly comparable with the
usual singularity theorems.

\section{Conclusions}

We have found that if a global time does not exist then it is
possible to infer that the spacetime is singular without making any
of the assumptions which are usually met in singularity theorems,
for instance that the cosmological flow is diverging everywhere in
the Universe, or that matter has clustered so much at some place in
the Universe that there exists a trapped surface.

Moreover, even if the spacetime started from a singularity, as it
can be inferred from Hawking's and Hawking and Penrose's theorems
\cite{hawking73}, it could still be  that this singularity be
related to the  {\em timelike} geodesic incompleteness of spacetime
which leaves open the possibility of applying the mentioned theorem
in a positive way to infer, starting from null completeness, the
existence of a time function. Actually, philosophically speaking,
null completeness is a very attractive requirement. Indeed, it is
possible to speculate that at early times the particles had no mass
(the Higgs field had not yet acquired a non-vanishing expectation
value), thus at early times the concept of observer as represented
by a timelike curve loses its meaning. Instead, only massless
particles make sense, and thus the concept of non-singularity must
be expressed from the `point of view' of massless particles.
Therefore the natural non-singularity requirement seems to be that
which demands a null geodesically complete Universe. Either used in
a `positive' or `negative' way, the theorem described in this work
gives new insights into the problem of the global existence of time.

\section*{Acknowledgments}
I  thank D. Canarutto for some suggestions that have improved the
readability of the manuscript. This work has been partially
supported by GNFM of INDAM.



\end{document}